\documentclass[11pt]{article}
\usepackage{amsmath}
\usepackage{amsthm}
\usepackage{amsfonts}
\def\CA{{\cal A}}
\def\CH{\hbox{$\cal H$}}

\def\C{\mathbb C}

\def\H{\mathbb H}

\def\tsa{\otimes_{\CA}}

\begin{document}
\title{Can (noncommutative) geometry accommodate leptoquarks?}
\author{ Mario Paschke \thanks{paschke@dipmza.physik.uni-mainz.de}, 
Florian Scheck\thanks{scheck@dipmza.physik.uni-mainz.de}, 
Andrzej Sitarz \thanks{sitarz@higgs.physik.uni-mainz.de},\\
\em Institut f\"ur Physik, \\ Johannes-Gutenberg Universit\"at, \\
 55099 Mainz, Germany }

\maketitle
\begin{flushright}
\vskip -7cm MZ-TH/97-31 \\
hep-th/yymmddd 
\end{flushright}
\vskip 7cm

\begin{abstract}
We investigate the geometric interpretation of the Standard Model
based on  noncommutative geometry. Neglecting the $S_0$-reality
symmetry one may introduce leptoquarks into the model. We give
a detailed discussion of the consequences (both for the Connes-Lott
and the spectral action) and compare the results with physical
bounds. Our result is that in either case one contradicts the 
experimental results.
\end{abstract}

\section{Introduction}

In the past years the Standard Model has been an object of
investigations directed towards its geometrical foundation within
the framework of noncommutative geometry (See \cite{Co,KIS} and the
references therein). The main idea
behind this concept is to generalize the notion of manifold and
differential structures to an algebraic setup and it appears that 
one may interpret the particle content of the Standard Model as
related to a {\em discrete noncommutative manifold}. Using this
input it is possible to derive the complete classical action, 
the weak hypercharges and all couplings between fermions and bosons. 
Moreover, the Higgs is naturally explained as a gauge boson related 
to the discrete differential structure, the symmetry breaking 
Higgs potential appearing as the Yang-Mills action for this gauge 
field. 

Some further speculations concern  mass relations \cite{KIS2} or the 
quantum group symmetry structure behind this model \cite{Co,Coq}. These 
could yield promising results, which might be easy to verify 
experimentally.

The reported anomaly in high-$Q$ $e^{\pm}p$ collisions at HERA
has aroused interest as a possible signal of physics beyond the
Standard Model \cite{HERA,Wil}. While these results have yet to be  
confirmed by other experiments some explanations have already been
proposed. Generally, it seems that within the models which are based
on a $SU(3)\times SU(2)\times U(1)$ gauge invariant Lagrangian
only scalar leptoquarks of certain type can explain the data \cite{Wil}.

In this letter we would like to discuss the predictions and 
constraints
on scalar leptoquarks which one gets from the noncommutative geometry 
description of the Standard Model. Let us note that contrary to some 
earlier results, terms which can break the $SU(3)$ symmetry are
admissible in the model, provided that one does not impose the 
so-called $S_0$ symmetry. We shall present the general construction
scheme and suggest what might prevent the breaking of the color 
symmetry even though leptoquarks are present.
         
Let us stress that the calculation of the differential calculus 
for a model of such complexity has not been considered before and is an
interesting topic in itself. Details are conveyed to the Appendix,
together with other technical observations concerning the model.

\section{The Standard Model in Noncommutative Geometry}

The crucial role in the model is played by the algebra $\CA = \C 
\oplus \H
\oplus M_3(\C)$ and its graded representation space $\CH$, a Hilbert 
space 
containing all particles of the Standard Model. The algebra acts on 
the elements 
of $\CH$ from the left and from the right, as shown in the following 
table :
\begin{center}
\begin{tabular}{|c||cc|c|c|}
\hline
 & ${\C}^*$ & $\C$ & $\H$ & $M_3(\C)$ \\ \hline \hline
${\C}^*$ & & $e_R$ &  & $d_R$ \\
$\C$ &  $\bar{e}_R$ & & $\bar{e}_L, \bar{\nu}_L$ & $u_R$ \\ \hline 
$\H$ & & $e_L, \nu_L$ & & $u_L, d_L$ \\ \hline
$M_3(\C)$ & $\bar{d}_R$ & $\bar{u}_R$ & $\bar{u}_L, \bar{d}_L$ & \\ 
\hline
\end{tabular}
\end{center}

Here, the convention is chosen such that the components of the 
algebra
act from the left along the rows and conjugated elements act from the 
right
along the columns. $\C^*$ means that the complex numbers act by 
multiplication with $\bar{z}$ instead of $z$. The left-handed 
particles are 
in doublets, on which quaternions act by their $2$-dimensional 
representation, 
and each quark has additional color indices on which $M_3(\C)$ acts.
 
The model has two symmetries, $\gamma$, which has values $+1$ for
right-handed particles and $-1$ for the left-handed, and an 
antilinear 
isometry $J$, which exchanges particles and antiparticles.

To construct the total Hilbert space of fundamental fermions one has
to take the tensor product with the bispinor bundle on the 
$4$-dimensional
manifold (for problems associated with the doubling of particles see 
\cite{Liz}). 

\section{Dirac operator and particle interactions.}

As in the case of gravity, the interactions between particles occur 
due to the 
presence of the generalized Dirac operator. The additional 
principle of gauge invariance requires the existence of bosonic gauge
fields, whose dynamics is set by the Yang-Mills action.

The coupling of the gauge boson fields is defined directly by the
structure of the Dirac operator. For the whole theory, it 
consists of two parts, the usual Dirac operator $ 
\gamma^\mu\partial_\mu$ 
on the $4$-dimensional manifold and the discrete Dirac operator $D_F$ 
which is a linear operator on $\CH$, satisfying certain symmetry 
restrictions. 

The gauge fields associated with the first one are the {\em usual}
vector bosons ($W^\pm,Z,\gamma,G^a$). The new phenomenon is the
appearance of bosons associated with the discrete part. 

Discrete noncommutative manifolds \cite{PaSi,Kr} allow only for
Dirac operators which link objects in the same row (in the 
same column) and connect particles of different chirality.

From the above table, we immediately get the possible actions of 
$D_F$,
between right-handed leptons (quarks) and the left-handed leptons 
(quarks),and of course, similarly, for the antiparticles:
\begin{eqnarray*}
D_F: & e_R \longleftrightarrow \left( e_L, \nu_L \right) \\ 
D_F: & u_R, d_R \longleftrightarrow \left( u_L, d _L \right)
\end{eqnarray*}

These are the only possibilities for the Dirac operator acting
between particles only (the conjugate would be among antiparticles).
 The principle of the $S_0$-reality condition \cite{Co,Kr} was, shortly 
speaking, to enforce that this is the case. Then there is no
direct coupling between leptons and quarks in the model and the
$SU(3)$ symmetry remains unbroken. The gauge potential induced
by this part is naturally interpreted as the Higgs field and one
can easily see that it couples to leptons and quarks as expected. 

However, it is easy to notice that if we do not require  
$S_0$-reality, the following chain of links is allowed:
$$D: \;\;  e_R \longleftrightarrow \left( e_L, \nu_L \right) 
\longleftrightarrow \bar{u}_R.$$

The existence of such a part of the Dirac operator has profound
consequences for the physical content of the model. First,
it allows for the existence of gauge bosons which couple
directly to left-handed leptons and the right-handed {\em up}
antiquark. Second, the resulting action possibly contains terms 
which break exact $SU(3)_c$, as feature which, of course, is 
unwanted. 
The new bosons would have the properties of scalar leptoquarks.

There is neither an  experimental nor a theoretical reason to exclude such 
particles from the model. Also, we have not found a compelling mathematical
(topological) advantage of this requirement.    

Before we present the action and discuss whether one can 
consistently include such leptoquarks, let us point out
what type of leptoquark is admissible. As shown above, the
model leaves room only for a scalar particle which couples
to right-handed $u$ antiquarks and the left-handed lepton doublet
(of course, there exists also the charge conjugated coupling).

Therefore, within this model one can make a strong 
prediction concerning the existence of {\em allowed} 
couplings, which could be tested experimentally.

\section{Construction of the action}

In this section we shall outline the calculation of the action
which one obtains for the leptoquarks. 
We shall not be interested in the couplings of fermions or other 
gauge 
bosons to leptoquarks, as they will not differ from the
usual gauge-invariant terms. Also the mass- or coupling constant
relations, which might possibly appear we leave for further
study. Our primary interest is to verify whether using the
general principle of Yang-Mills theory one may obtain a
consistent model without $SU(3)$ symmetry breaking. 
Regarding more technical details we refer the reader
to the appendix.

Models based on noncommutative geometry usually are constructed
using the Connes-Lott action principle \cite{CoLo} for gauge theories, 
which
is the noncommutative extension of the Yang-Mills action. More 
recently
a {\em spectral action principle} was proposed \cite{CoCh}. While for
classical differential manifolds they yield the same result, they 
differ
significantly when one includes the noncommutative discrete structure. 
Here
we shall briefly sketch both approaches.

\subsection{Connes-Lott Action}

The main steps in constructing  the Connes-Lott action is 
the determination of the differential structure and the scalar 
product, 
in particular for the bimodule of two-forms. 
This is usually done as follows, first an algebra
$\Omega_D(\CA)$ is constructed, as the subalgebra of operators 
generated by
$\CA$ and commutators $[D,a]$ for $a \in \CA$. In a natural way it is 
an
image of the {\em universal differential algebra}, $\Omega_u(\CA)$, 
which
consists of elements $a_0 da_1 \ldots da_k$, $a_i \in \CA$, with the
differential map $d$, $ d (a_0 da_1 \ldots da_k) = da_0 da_1 \ldots 
da_k $,
satisfying the usual Leibniz rule and $d^2=0$.  Now, using the map 
$\pi :
\Omega_u(\CA) \to \Omega_D(\CA)$ one can find a differential algebra
$\Omega(\CA)$ such that the kernel of the differential map $\pi_d:
\Omega_u(\CA) \to \Omega(\CA)$ contains the kernel of the map $\pi$.  
The
construction is unique if one postulates that the obtained 
differential
algebra $\Omega(\CA)$ is maximal. Since the algebra $\Omega_D(\CA)$ 
is
equipped with a natural scalar product (as a subalgebra of the 
operator
algebra), one has only to choose an appropriate embedding of 
$\Omega(\CA)$
in $\Omega_D(\CA)$, this is usually done as an orthogonal embedding, 
i.e.,
the image of a form must be orthogonal to the kernel of the 
projection on
differential forms. For details see \cite{KIS}.

As we are interested here in qualitative answers, we
restrict ourselves only to some part of the Hilbert space and
Dirac operator,  which describes the leptons
and the right-handed antiquark $\bar{u}$. Here both the Higgs 
and leptoquark sector play a role. What we leave out is the quark 
sector where only the standard Higgs appears.

The algebra $\CA =\C \oplus \H \oplus M_3(\C)$ acts on the
chosen sector of the Hilbert space $\CH =\C \oplus \C^2 \oplus \C^3$, 
as $\bar{z} \oplus q \oplus m$, $z\in\C, q \in \H, m \in M_3(\C)$. 

The allowed Dirac operator is represented as :

$$ \left( \begin{array}{ccc} 0 & a & 0 \\ a^\dagger & 0 & b \\
0 & b^\dagger & 0 \end{array} \right), $$
where $a : \C^2 \to \C$ and $b: \C^3 \to \C^2$ are a priori 
arbitrary (complex) linear operators. We can use our knowledge
of the Standard Model to associate the masses of the leptons with
$a$. 

The gauge bosons related with the discrete differential structure
appear to be represented by a doublet $\Phi$ of complex fields 
(the Higgs) and six (a doublet whose components are triplets with 
regard to 
$SU(3)$) complex fields, which we shall call $\Psi$, having the 
following gauge transformation rules:
$$ \Phi' = U_1 \Phi U^\dagger_2, \;\;\;\; \Psi' = U_2 \Psi U^\dagger_3,$$ 
where $U_1,U_2,U_3$ denote, respectively, $U(1), SU(2), SU(3)$
transformations.

Details of the action calculation are given in the Appendix. In 
general, 
there will be three terms which contribute under certain 
circumstances
 to the action:

\begin{enumerate}
\item Higgs self-interaction term:
$$ a \left( \Phi \Phi^\dagger - 1  \right)^2 a^\dagger, $$ 
\item Higgs and leptoquark self-interaction term:
$$ \sim {\rm Tr}\left( \Phi^\dagger a^\dagger a \Phi - a^\dagger a
+ \Psi \Psi^\dagger - b b^\dagger \right)^2, $$
\item Higgs - leptoquarks coupling:
$$ a \left( \Phi \Psi \Psi^\dagger \Phi^\dagger \right) a^\dagger.$$
\end{enumerate}

The first one is the well-known symmetry breaking Higgs potential,
the second one generates both $SU(3)$ symmetry breaking 
as well as mass terms for Higgs and leptoquarks, whereas the third
one (which occurs only under the condition $ab=0$) gives a contribution
to the mass-terms for leptoquarks.

Of course, physically,  the most interesting situation would be to 
have no color symmetry breaking but massive leptoquarks. We shall
briefly discuss all possibilities within the model and slight 
extensions
thereof:

\begin{itemize}

\item If the potential term exists there is $SU(3)$ spontaneous
symmetry breaking. In this situation the gluons become massive. 
Moreover,
there will be direct interaction terms between leptons
and quarks (arising from the vacuum expectation value of $\Psi$),
which lead to lepton and baryon number violation. Note that since
$\Psi$ carries lepton and baryon numbers the vacuum would also have
such quantum numbers. Such a model is phenomenologically 
unacceptable.

\item If the leptoquarks couple {\em diagonally} to all families and,
moreover, the couplings are identical, then the leptoquark part of 
the 
curvature form, i.e. the second term, vanishes.  However, in such a situation 
either both
leptoquarks or at least one component must be massless. This seems to be
excluded experimentally because massless leptoquarks should be observable
in atomic spectroscopy.

\item If the differential structure on the algebra {\em does not} 
coincide with the one introduced\footnote{One may, for instance, 
argue
that it comes from the quantum group structure related with the 
algebra \cite{PSforth}.}
it is well possible that even without the condition on diagonality, 
we 
would have no potential term. The problem with the masses remains.

\item We should mention that the extension of the model including massive 
neutrinos
\footnote{Note that here Poincar\'e duality would not be 
satisfied!}, 
would not solve the mass problem. If one still assumes a diagonal coupling
of the leptoquark to the families, it would decouple from the Higgs and thus
remain massless. ( The condition $ab=0$ will not be fulfilled in this 
situation with $b \neq 0$.)
\end{itemize}

\subsection{The spectral action}

We devote a separate paragraph to the action obtained using the 
spectral
principle of Connes \cite{CoCh,Co}. Here, the action is related to the 
eigenvalues
of $D^2$, $D$ being the Dirac operator. Such an action, applied to 
the
tensor product of classical geometry with the considered 
noncommutative
manifold  yields both Yang-Mills and gravity terms. Although the 
principle
includes a large number of free parameters and although one 
additionally obtains 
some unrealistic higher-order terms for gravity, the Yang-Mills-Higgs 
action
for the Standard Model and fermionic actions are recovered correctly. 
However, the whole picture is changed dramatically because the discrete 
differential structure becomes irrelevant and,\footnote{In fact, only 
$\Omega^1$
plays a role.} for instance, one may not speak any longer of the 
Higgs potential
term as arising from the curvature of the Higgs connection.

Nevertheless it is worth investigating what the action including the 
leptoquarks
would look like. Shortly speaking one obtains similar terms as in the 
Connes-Lott
model, the difference being, however, in the much smaller number of free 
parameters.
The most important difference comes , of course, from the fact that one does not
use $\Omega^2$:  one can not 
get rid of the potential terms for the leptoquark by assuming a diagonal 
coupling to the families.
Generally
one gets Higgs and leptoquark self-interaction terms as well as
a Higgs-leptoquark interaction term.   

The resulting potential can be written as
\begin{eqnarray*} 
V(\vec{\psi},\phi) 
& = & \alpha \left( |{\vec \psi}_1|^4 + |{\vec \psi}_2|^4+ 
     2 |{\vec \psi}_1 \cdot {\vec \psi}_2 |^2 \right)
 -\frac{\mu^2}{2} \left( |{\vec \psi}_1|^2 + |{\vec \psi}_2|^2 \right) \\
 & & + \delta |{\vec \psi}_1|^2\phi^2 
     + \lambda \phi^4 - \frac{\mu^2}{2} \phi^2.
\end{eqnarray*}
where all parameters are positive and $\alpha > \lambda$.

Note that ${\vec \psi}_2 $ decouples from the Higgs $\phi$. Moreover, $\mu^2$ 
being positive, there is no mass term for this field if we assume that the 
vacuum expectation value of ${\vec \psi}_1 $ vanishes.
Even more so, the  minimum of this potential clearly requires  
${\vec \psi}_2$ and ${\vec \psi}_1 $ to be orthogonal. But then ${\vec \psi}_2$
decouples from all other fields. The minimum for its potential is provided
if \[ |{\vec \psi}_2|^2 = \frac{\mu^2}{4\alpha}, \]
it breaks $SU(3)$ spontaneously and violates lepton and baryon number 
conservation. \\
The vacuum expectation value of ${\vec \psi}_1 $ is zero, while  
$\phi_0^2 = \frac{\mu^2}{4\lambda}$ as usual. 
Additionally, although less catastrophic, it turns out that one can not adjust
the couplings of the leptoquark to the fermions to be consistent with the 
present experimental bounds. 
                                        
\section{Conclusions}

The main conclusion of this Letter is that it is possible to 
accommodate
leptoquarks in the usually assumed model based on noncommutative 
geometry. 
The price one pays is the breaking of the so-called
$S_0$-reality condition.

Taking such a model as input one obtains a stringent prediction 
concerning the possible type of leptoquarks: if they exist only
couplings between left-handed leptons and a right-handed
$\bar{u}$ antiquark can exist. 
As a
consequence there cannot be any anomaly in $e^-p \to eX$
events at HERA. Moreover, the missing evidence of a charged current
signal $e^+p \to \nu X$ favors scalar leptoquarks. 
As far as other bounds are concerned this
type of leptoquark seems not to be excluded experimentally.

However, the construction of the action leads to several problems and 
it 
is difficult to construct a consistent model without the breaking of 
color 
symmetry or getting massless leptoquark states.
The problem with breaking color symmetry in the model can be dealt 
with if one uses the Connes-Lott action. 
Bounds on flavor mixing which suggest that leptoquarks couple almost 
diagonally to families, do not contradict the model. On the contrary,
diagonal or almost diagonal coupling seems to be significant for
the strong symmetry to remain unbroken. \\
 The main problem are the leptoquark masses, atomic spectroscopy 
seems to exclude massless leptoquarks, moreover, a bound arising from the
$\rho$-parameter suggests that the two isospin components be nearly 
mass degenerate. 
Both, the Connes-Lott action as well as the spectral action 
contradict this experimental data. Of course, one cannot exclude that 
the whole 
principle of constructing the action must be modified in which case
the results might change dramatically.\\

On the other hand, it might be possible to find a theoretical reason that
excludes the appearance of leptoquarks. One could, for instance, 
{\em require} the mentioned $S_{0}$-reality. 
Another idea, which we find more attractive, is to enforce the absence of the
coresponding part of the Dirac-operator
by a {\em principle}, which is directly related to the 
structure of differential calculus and symmetries \cite{PaSi,PSforth}.\\

\vspace{1cm}
\begin{appendix}
\begin{center}
{\Large \bf Appendix}
\end{center}
\vspace{.5cm}
We discuss here the technical details of the model construction.
Before we turn to the noncommutative differential calculus it is instructive
to examine the gauge invariant terms, which can possibly
appear in the potential for the Higgs and the leptoquarks.
Clearly, only the couplings of the two types of scalar fields are interesting.
Let us denote the six components of the scalar field $\Psi$ as 
$ \psi^{i}_{k}$ $ i=1,\ldots,3$ $ k=1,2 $
 and the two components of the Higgs $\Phi$ as
$ \phi _{k}$ $ k=1,2 . $\\
Note that, due to the freedom of a $SU(2)$-gauge transformation, we can assume
\[ \phi_{2}=0,\qquad \qquad\phi=\phi_{1} \;\;{\rm real} .\]
There are two different possibilities:
\begin{itemize}
\item   
\[ \Phi^{\dagger} \Phi \; {\rm Tr}\left( \Psi^{\dagger} \Psi \right) =
 \phi^{2} \left( \left| {\vec \psi_{1}}\right|^{2} +\left| 
{\vec \psi_{2}}\right|^{2} \right)\]
In this case the vacuum expectation value of the Higgs will lead to a 
mass term for all six components of the leptoquark. The masses are
degenerate. Unfortunately, this term will not appear in the models, which
are based on noncommutative geometry.
\item
\[  \Phi^{\dagger}\Psi \Psi^\dagger \Phi  =
 {\rm Tr}\left( \Psi^{\dagger}\Phi \Phi^\dagger \Psi \right) =
  {\rm Tr}\left( \Psi \Psi^{\dagger} \Phi\Phi^\dagger \right) =
  \phi^{2}  \left| \vec{\psi}_1\right|^{2} \]
Here the mass terms for the components $\vec{\psi}_{2}$ will not get
a contribution from the Higgs' expectation value.
\end{itemize}
\section{The discrete differential structure}

Given the representation and the Dirac operator we can now construct 
the differential algebra, following the usual procedure \cite{KIS2,PaSi}.

$\Omega^1(\CA)$ is isomorphic to a bimodule of operators on $\CH$
of the following form:

$$ \left( \begin{array}{ccc} 0 & \bullet & 0 \\ \bullet & 0 & \bullet 
\\
0 & \bullet & 0 \end{array} \right), $$

where the possible nonvanishing entries (bullets) can be arbitrary. 
Left-
and right multiplication by the elements of $\CA$ is the usual matrix
multiplication.

$\Omega^2(\CA)$ is a quotient of $\Omega^1(\CA) \tsa \Omega^1(\CA)$
by the subbimodule  generated by the commutators $[r,D^2]$ 
for $r \in \CA$.

It is clear that $\Omega^1(\CA) \tsa \Omega^1(\CA)$ is represented on
$\CH$ as operators:

$$ \left( \begin{array}{ccc}  \bullet & 0 & \bullet \\ 0 & \bullet & 
0  \\
\bullet & 0 & \bullet  \end{array} \right), $$

again with arbitrary entries.

The interesting part is the subbimodule that we have to quotient out. 
$D^2$ becomes:

$$ \left( \begin{array}{ccc}  aa^\dagger & 0 &  ab \\ 0 & a^\dagger a 
+ b b^\dagger  & 0  \\
b^\dagger a^\dagger & 0 & b^\dagger b  \end{array} \right), $$

It is clear that we have to consider two separate situations:

\begin{itemize}

\item $ab=0$ and $a\not=0$ and $b\not=0$ \footnote{Of course, 
$a\not=0$ as we
know from the phenomenology of the SM, if $b=0$ then the whole
discussion reduces to the previously widely discussed models [!]}

\item  $ab\not=0$

\end{itemize}

We begin with the former.

For $ab=0$ we have to distinguish two cases, depending on the value
of $a^\dagger a + b b^\dagger$. Before we do so, let us make a remark
concerning decomposition of $M_2(\C)$ as a bimodule of $\H$.

{\bf Remark:} $M_2(\C) \sim \H \oplus \H$ as a bimodule over $\H$. 
Unless an element of $M_2(\C)$ belongs to one of these components
of the direct sum it generates the whole of $M_2(\C)$.
 
\begin{itemize}

\item $a^\dagger a + b b^\dagger$ is proportional to $1$

Then, this part of $D^2$ commutes with everything and therefore
the bimodule  generated by $[r,D^2]$ is of the form:
$$ \left( \begin{array}{ccc} 0 & 0 & 0 \\ 0 & 0 & 0  \\
0 & 0 & M_3(\C) \end{array} \right). $$

\item $a^\dagger a + b b^\dagger$ is not proportional to $1$.

Then only the traceless part of $a^\dagger a + b b^\dagger$ 
contributes to the bimodule of the junk, its most general form
is:

$$ \left( \begin{array}{cc} r & z  \\ \bar{z} & -r \end{array} 
\right), $$

and one can verify that it is of the form $q i$, where $q$ is a 
quaternion
given by the pair $(r, z)$ and $i$ is the matrix $\hbox{diag}(1,-1)$, 
which
generates the subbimodule isomorphic to $\H$ in $M_2(\C)$. Therefore,
the ideal generated by $[a,D^2]$ looks like:

$$ \left( \begin{array}{ccc} 0 & 0 & 0 \\ 0 & \H & 0  \\
0 & 0 & M_3(\C) \end{array} \right). $$ 

\end{itemize}

If $ab \not=0$, the whole procedure is quite similar, however,
additionally we have to take care of the off-diagonal entries of
$D^2$.  Remember \cite{PaSi} that apart from the bimodule generated
by $[a,D^2]$, we would have contributions from elements of the
form $\sum_i a_i(\Xi - \xi\xi)b_i$ for any $a,b$ such that 
$a_i \xi b_i = 0$. For us, $\Xi-\xi\xi$ is the part of $D^2$ with
off-diagonal elements only, and it is easy to verify that it
suffices to take $a$ from $\C$  and $b$ from $M_3(\C)$ to satisfy
the requirement $a\xi b=0$, however then, one can generate 
elements of the bimodule with following (arbitrary) entries:

$$ \left( \begin{array}{ccc} 0 & 0 & \bullet \\ 0 & 0 & 0  \\
\bullet & 0 & 0 \end{array} \right). $$

Therefore we can now ignore the off-diagonal entries, but for
the diagonal part we have already established what kind of bimodule
is generated by them (all considerations were independent of
the $ab=0$ condition).

\subsection{Differential algebra for the tensor product with the 
manifold}

A surprising feature of spectral triple noncommutative geometry is 
that
tensoring two spectral triples can change the differential structure 
on 
the components. Here, the representation image of $\Omega^1(\CA) 
\otimes_\CA \Omega^1(\CA)$ on the Hilbert space intersects the image of
the $\Omega^1(M) \otimes_\CA \Omega^1(M)$ for any $4$-dimensional 
manifold $M$.  Therefore the differential ideal one has to quotient 
in order
to obtain $\Omega^2$ for the tensor product gets enlarged. Its 
restriction
to the discrete component becomes just the image of the algebra $\CA$ 
itself. 

\subsection{Scalar fields, gauge theory and the Connes-Lott action}

We calculate here the discrete part of the gauge curvature and the total action 
for the Connes-Lott model.
For the calculation of $dA$ and $AA$ we might restrict ourselves to 
the $ab=0$ case, 
as in the case $ab \not=0$ the possible additional terms would be 
off-diagonal and - as we 
already know, they would be in the subbimodule which one divides out.

The gauge potential is a self-adjoint one-form $A$, which we shall 
parametrize in the 
following way.

$$ A  = \left( \begin{array}{ccc} 0 & a(\Phi-1) & 0 \\ 
(\Phi^*-1)a^\dagger & 0 &  \Psi-b \\
0 & \Psi^\dagger -b^\dagger& 0 
\end{array} \right), $$

where $\Psi$ is a $2 \times 3$ matrix and $\Phi$ is a quaternion. The 
shift in the parametrization
is to simplify the formulas and use physical fields $\Phi$ and $\Psi$ 
which transform
homogeneously under gauge transformations.

Then using \cite{PaSi} we find:

$$ dA = \left( \begin{array}{ccc}
a(\Phi - 1)a^\dagger  &          0      &  a(\Psi - b) +a(\Phi - 1)b  
\\
        0     & a^\dagger a (\Phi - 1) + b (\Psi - b)^\dagger & 0 \\
         0    & 0  & b^\dagger (\Psi - b)  
\end{array} \right) + \hbox{h.c.} $$

and for $AA$:
$$ AA = \left( \begin{array}{ccc}
a (\Phi - 1)^\dagger(\Phi - 1)a^\dagger & 0 & a(\Phi - 1)(\Psi - b) 
\\
0 & (\Phi - 1)^* a^\dagger a (\Phi - 1) + (\Psi - b) (\Psi - 
b)^\dagger   & 0 \\
(\Psi - b)^\dagger (\Phi - 1)^* a^\dagger 
& 0 & (\Psi - b)^\dagger (\Psi - b)  \end{array} \right),$$

Taking into account the form of $\Omega_2$ for various situations
we may now write the resulting Yang-Mills action $S= (F,F)$ for 
various 
situations:

\begin{itemize}

\item $a^\dagger a + b b^\dagger$ proportional to $1$

\begin{itemize}
  \item $ab=0$

$$ S = a ( |\Phi|^2-1 ) a^\dagger + a (\Psi\Phi) (\Psi\Phi)^\dagger 
a^\dagger
+ {\rm Tr} \left| \Phi^* a^\dagger a \Phi - a^\dagger a + 
\Psi \Psi^\dagger - bb^\dagger\right|^2. $$

  \item $ab\not=0$

$$ S = a ( |\Phi|^2-1 )a^\dagger + 
 {\rm Tr}\left| \Phi^* a^\dagger a \Phi - a^\dagger a + \Psi \Psi^\dagger - 
bb^\dagger \right|^2. $$
\end{itemize}

\item $a^\dagger a + b b^\dagger$ not proportional to $1$ 

\begin{itemize}
  \item $ab=0$

$$ S = a ( |\Phi|^2-1 ) a^\dagger + a (\Psi\Phi) (\Psi\Phi)^\dagger 
a^\dagger
+\left| \frac{1}{2} \hbox{Tr}  \left(\Phi^* a^\dagger a \Phi - 
a^\dagger a + 
\Psi \Psi^\dagger - bb^\dagger\right)\right|^2. $$
  
  \item $ab\not=0$

$$ S = a ( |\Phi|^2-1 ) a^\dagger  
+\left| \frac{1}{2} \hbox{Tr}  \left(\Phi^* a^\dagger a \Phi - 
a^\dagger a + 
\Psi \Psi^\dagger - bb^\dagger\right)\right|^2. $$
\end{itemize}
\end{itemize}

Let us briefly describe the minima for the various situations.
First, in all cases the action is a sum of positive terms. Thus we have 
to find the solutions of $S=0$.\\ 
If $ab\neq 0$ the Higgs potential term $a ( |\Phi|^2-1 ) a^\dagger$ requires
$ \Phi$ being unitary, by a gauge we can take $\Phi=1$. The remaining
term reduces to $\Psi \Psi^\dagger = bb^\dagger$ i.e $\Psi=bU^\dagger$, where 
$U\in U(3)$. Clearly this solution leads to spontaneous breaking of the
color symmetry.\\ 
For $ab=0$ the solution is the same, since then (if $\Phi=1$, 
$\Psi=b$) the term $a\left(\Psi\Phi\right)$ vanishes.
This is a consequence of a theorem that has been proven in \cite{PaSi}.

\subsection{Physical fields and spontaneous symmetry breaking} 
The action that we have calculated in the previous subsection is still
preliminary, since only the discrete differential structure has been used.
Taking into account the complete differential algebra, one has to take care
about the enlarged differential ideal.   

Recall that now we also have to quotient out the subbimodule,which is 
isomorphic to the algebra itself. If there were only one family of fermions,
this would lead to a vanishing potential for the scalar fields. Adding 
families,
we can assume that $b^\dagger b$ is of the form $B\otimes 
\hbox{id}_{N_{f}}$. In other words, we assume that the leptoquarks 
couple diagonally to the fermion
families. Phenomenologically this requirement is very attractive \cite{Wil}.
In this situation the $\Psi\Psi^\dagger$, $bb^\dagger$ are in the
algebra and the potential reduces to
\[ V(\Psi,\Phi)= V(\Phi)  + a (\Psi\Phi) (\Psi\Phi)^\dagger 
a^\dagger , \]
(if $ab=0$) and there is no symmetry breaking self-interaction term for $\Psi$.

Unfortunately, there is also no mass term for $\Psi$, except the one that 
comes from the expectation value of $\Phi$. Thus, at least three components of 
the field $\Psi$ will remain massless 
\footnote{If neutrinos are massive, $a$ will be a $2\times 2$ matrix
with eigenvalues $m_{e},m_{\nu}$. In this case all the components of $\Psi$
would, at first sight, obtain masses from the coupling with the Higgs. However,
since the rank of $a$ will then be 2, there will be no nontrivial solution of
$ab=0$. Thus the leptoquarks and the Higgs will decouple completely.}.
  
\section{The spectral action}

The new {\em spectral principle} relates the bosonic part of the action 
with the eigenvalues of the square of the Dirac operator. It is
defined by:
\[  S_{B} = {\rm Tr}\chi\left(\frac{{\cal D}^2}{\Lambda^2} \right), \]
and can easily be computed using the heat kernel technique \cite{Gilkey}:
\begin{equation}
S_{B}= \frac{1}{16 \pi^2}\int_{M}\left(\Lambda^4 f_{0}a_{0} + \Lambda^2 
f_{2}a_{2} + f_{4}a_{4} + \Lambda^{-2} f_{6}a_{6} + \ldots \right) {\rm d}v(x).
\end{equation}
Here $f_{i}$ are the usual moments of the function $\chi$, and
the first three nonvanishing heat kernel coefficients $a_{n}(x,P)$ can 
directly be written down \cite{Gilkey} if one casts $P$ in the form:
\begin{equation}
P:= \frac{{\cal D}^2}{\Lambda^2} = - \left( g^{\mu\nu} 
    \partial_{\mu}\partial_{\nu}
    + A^\mu \partial_{\mu} + B \right)
\end{equation}
We shall only retain the contributions from $a_{0},a_{2}$ and $a_{4}$.
Since the calculation is straightforward, we shall not give the complete 
expression for $P$ and we shall only state the resulting bosonic action. 

Before we do so, it is necessary to comment on the free parameters of the
theory. They are $\Lambda$, $f_{0}$, $f_{2}$ and $f_{4}$. In the situation with
$b=0$, the operator ${\cal D}^2$ has a block diagonal form, acting on
each of the three lepton families and the quarks separately. This offers the
possibility to introduce four further parameters $x$, $y_{i}$ by modifying the
definition of the bosonic action as  
\[ S_{B} = x{\rm Tr}_{Q} \chi \left(  \frac{{\cal D}^2}{\Lambda^2} \right) +
           \sum_{i=1}^{3} y_{i}
          {\rm Tr}_{i} \chi \left(  \frac{{\cal D}^2}{\Lambda^2} \right),
\]
where ${\rm Tr}_{i}$ denotes the trace in the subspace spanned by the
i-th lepton family, and ${\rm Tr}_{Q}$ is the trace over the subspace spanned
by the quarks \footnote{
Note that it is also possible to take four different functions
$\chi_{i},\chi_{Q}$}.

In our case with $b\neq 0$, there is no such decomposition of ${\cal H}$ since
${\cal D}^2$ mixes leptons and antiquarks. The only free parameters are 
therefore $\Lambda, f_{i}$. Additionally, we have still not identified the 
parameters of $b$ with physical quantities. Let us assume that $b$ is of
the form 
\[ b= {\tilde b}\otimes {\rm diag}(\kappa_{e},\kappa_{\mu},\kappa_{\tau}). \]
Then the leptoquark couples diagonally to the families, with coupling constants
\[ k_{i} \sim \frac{\kappa_{i}}{\sqrt{\sum_{i} f_4\kappa_{i}^2}}. \]

We shall write the components of the leptoquark as vectors ${\vec \psi}$.
The resulting potential for the scalar fields (in the limit of 
flat spacetime) is then given as:
\begin{eqnarray*} 
V(\vec{\psi}_1,\vec{\psi}_2,\phi) & = & 
\alpha \left( |{\vec \psi}_1|^4 + |{\vec \psi}_2|^4+ 
     2 |{\vec \psi}_1 \cdot {\vec \psi}_2 |^2 \right)
 -\frac{\mu^2}{2} \left( |{\vec \psi}_1|^2 + |{\vec \psi}_2|^2 \right) \\
& &  +\delta |{\vec \psi}_1|^2\phi^2 
     +\lambda \phi^4 - \frac{\mu^2}{2} \phi^2.
\end{eqnarray*}
The parameters of this potential are explicitly given as
\begin{eqnarray*}
\alpha & = & \frac{2\pi^2 K_2}{f_{4} K^2},\\
\mu & = & 2 \frac{f_2}{f_4}\Lambda^2, \\
\delta & = & \frac{2\pi^2 M}{f_{4} KL},\\
\lambda & = & \frac{\pi^2 L_2}{f_{4} L^2},
\end{eqnarray*}
where 
\begin{eqnarray*} 
K_2 & = & \sum_{i} \kappa_{i}^4,\\
K & = & \sum_{i} \kappa_{i}^2,\\
L & = &  \sum_{i} (3m_{u_{i}}^2 + 3m_{d_{i}}^2 +m_{e_{i}}^2),\\
L_2 & = &  \sum_{i} (3m_{u_{i}}^4 + 3m_{d_{i}}^4 +m_{e_{i}}^4), \\
M & = &  \sum_{i} \kappa_{i}^2(m_{u_{i}}^2 + m_{d_{i}}^2 +m_{e_{i}}^2),
\end{eqnarray*}
This potential leads to a nonvanishing vacuum expectation value of 
$|{\vec \psi}_2|^2$ and thus to spontaneous breaking of color symmetry.
\end{appendix}

\end{document}